\documentclass[twocolumn,floatfix,aps,superscriptaddress,nofootinbib]{revtex4-2}
\usepackage{graphicx}
\usepackage{amsmath}
\usepackage{amssymb}
\usepackage{bm}
\usepackage{hyperref}

\begin{document}

\title{Reflectionless Klein tunneling of Dirac fermions: Comparison of split-operator and staggered-lattice discretization of the Dirac equation}
\author{A. Don\'{i}s Vela}
\affiliation{Instituut-Lorentz, Universiteit Leiden, P.O. Box 9506, 2300 RA Leiden, The Netherlands}
\author{G. Lemut}
\affiliation{Instituut-Lorentz, Universiteit Leiden, P.O. Box 9506, 2300 RA Leiden, The Netherlands}
\author{M. J. Pacholski}
\affiliation{Instituut-Lorentz, Universiteit Leiden, P.O. Box 9506, 2300 RA Leiden, The Netherlands}
\author{J. Tworzyd{\l}o}
\affiliation{Faculty of Physics, University of Warsaw, ul.\ Pasteura 5, 02--093 Warszawa, Poland}
\author{C. W. J. Beenakker}
\affiliation{Instituut-Lorentz, Universiteit Leiden, P.O. Box 9506, 2300 RA Leiden, The Netherlands}
\date{April 2022}
\begin{abstract}
Massless Dirac fermions in an electric field propagate along the field lines without backscattering, due to the combination of spin-momentum locking and spin conservation. This phenomenon, known as ``Klein tunneling'', may be lost if the Dirac equation is discretized in space and time, because of scattering between multiple Dirac cones in the Brillouin zone. To avoid this, a staggered space-time lattice discretization has been developed in the literature, with \textit{one} single Dirac cone in the Brillouin zone of the original square lattice. Here we show that the staggering doubles the size of the Brillouin zone, which actually contains \textit{two} Dirac cones. We find that this fermion doubling causes a spurious breakdown of Klein tunneling, which can be avoided by an alternative single-cone discretization scheme based on a split-operator approach.
\end{abstract}
\maketitle

\section{Introduction}
\label{sec_intro}

Massless Dirac fermions have an energy-independent velocity, so if they move uphill in a potential landscape they are not slowed down. Even an infinitely high potential barrier cannot stop a particle approaching along a field line. This counterintuitive behavior is referred to as the Klein paradox, and the perfect transmission through a potential barrier is called Klein tunneling. It plays a central role in the ``electron quantum optics'' of Dirac materials, such as graphene,  topological insulators, and Weyl semimetals \cite{All11,Bee08}.

The Dirac fermions on the two-dimensional (2D) surface of a 3D topological insulator are of particular interest because they work around the ``no-go'' theorem for the impossibility to place a single species of massless Dirac fermions on a lattice \cite{Nie81}. The work-around consists in spatially separating two Dirac cones, one on the top surface and one on the bottom surface of the insulating material \cite{Vaf14,Kim15}. An unpaired Dirac cone is topologically protected: electrostatic disorder cannot open up a gap and Klein tunneling is fully reflectionless.

Computer simulations of the electron dynamics on the 2D surface could work with a 3D lattice, but because this is computationally expensive there is a need for methods to implement a single Dirac cone on a 2D lattice \cite{Tong}. Here we compare two such methods, using Klein tunneling as a test case for the presence or absence of fermion doubling. 

Both methods discretize the time-dependent Dirac equation, 
\begin{equation}
i\hbar\frac{\partial}{\partial t}\Psi(\bm{r},t)=v_0\sum_{\alpha=x,y} (p_\alpha+eA_\alpha)\sigma_\alpha\Psi(\bm{r},t)+V\Psi(\bm{r},t),\label{Diraceq}
\end{equation}
where $v_0$ is the energy-independent velocity of the massless electrons (Dirac fermions), $V$ and $\bm{A}$ are scalar and vector potentials, and the $\sigma_\alpha$'s are Pauli spin matrices. One method works in real space on a staggered space-time lattice \cite{Ham14,Pot17a}, the other method works in Fourier space using a split-operator technique \cite{Don22}.

The staggered-lattice discretization is due to Hammer, P\"{o}tz, and Arnold (HPA) \cite{Ham14,Pot17a}, and has been applied to a variety of problems in condensed matter physics \cite{Ham13,Pot16,Pot17b,Pot21}. For free fermions ($V,\bm{A}\equiv 0$) it has the bandstructure
\begin{equation}
\sin^2(\varepsilon\delta t/2)=\gamma^2\sum_{\alpha=x,y} \sin^2(a_0k_\alpha/2),\;\;\gamma\equiv \frac{v_0\delta t}{a_0}\leq \frac{1}{\sqrt 2}.\label{dispersionsin}
\end{equation}
Here $a_0$ and $\delta t$ are the lattice constants in space and time; $\bm{k}$ and $\varepsilon$ are crystal momentum and quasi-energy.\footnote{The quasi-energy $\varepsilon$ is such that $\Psi(t+\delta t)=e^{i\varepsilon \delta t}\Psi(t)$, so the quasi-energy spectrum repeats itself with period $2\pi/\delta t$.}

The split-operator discretization \cite{Don22} builds on early work of Stacey \cite{Sta82,Two08,Lem21}. The bandstructure has the same form as Eq.\ \eqref{dispersionsin} --- but with the sine replaced by a tangent,\footnote{The tangent $\tan(a_0k_\alpha/2)$ has a pole at the Brillouin zone boundary $k_\alpha=\pm\pi/a_0$, but the pole cancels from Eq.\ \eqref{tangentdispersion}, which has a continuous quasi-energy dispersion $\varepsilon(\bm{k})$ for any real $\gamma$.}
\begin{equation}
\tan^2(\varepsilon\delta t/2)=\gamma^2\sum_{\alpha=x,y}\tan^2(a_0k_\alpha/2).\label{tangentdispersion}\end{equation}

A  unique property of the HPA technique is that it is fully gauge invariant \cite{Ham14,Pot17a}. It is also highly efficient, because the time evolution is governed by a direct, rather than implicit, difference equation, which moreover is local in real space. These features are lacking in the split-operator discretization \cite{Don22}, which motivated us to compare the two approaches in some detail.

Our central finding, presented in Sec.\ \ref{sec_BSdoubling}, is that the bandstructure \eqref{dispersionsin} from the staggered-lattice discretization actually has two inequivalent Dirac cones in the first Brillouin zone: The Dirac points at $\bm{k}=0$ and $\bm{k}=(2\pi/a_0,0)$ are not related by a reciprocal lattice vector. This Brillouin zone doubling is avoided in the split-operator discretization. We assess the consequences for Klein tunneling in Sec.\ \ref{sec_consequence} and conclude in Sec.\ \ref{sec_conclude}.

\section{Brillouin zone doubling}
\label{sec_BSdoubling}

\begin{figure}[tb]
\centerline{\includegraphics[width=0.9\linewidth]{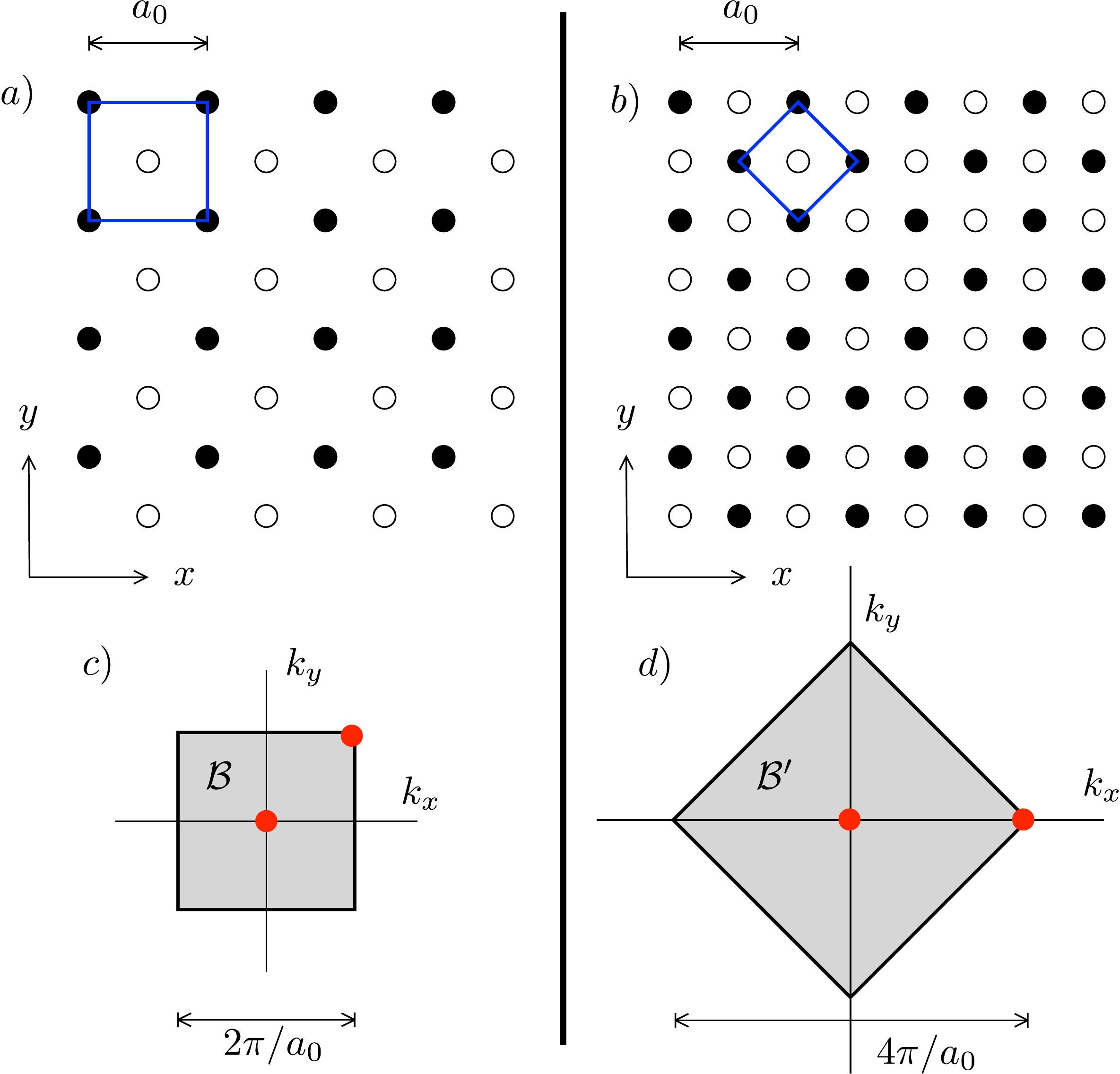}}
\caption{Comparison of two types of staggered grids for the spatial discretization of Dirac fermions, in the Susskind fermion approach (panel a, corresponding Brillouin zone ${\cal B}$ in panel c) and in the HPA modification (panel b, Brillouin zone ${\cal B}'$ in panel d). The black and white dots distinguish the $u$ and $v$ amplitudes of the spinor wave function $\Psi=(u,v)$. The blue squares give the unit cell of the lattice in real space, the grey square is the first Brillouin zone in momentum space, the red dots indicate two inequivalent Dirac points.
}
\label{fig_grids}
\end{figure}

The HPA technique modifies a staggered lattice discretization known as Susskind fermions \cite{Kog75,Sus77} and implemented in $2+1$ space-time dimensions in Ref.\ \onlinecite{Ham14a}. In that approach the two components of the spinor $\Psi=(u,v)$ are discretized on separate lattices, displaced (staggered) from each other by $a_0/2$ and evaluated at alternating time slices (see Fig.\ \ref{fig_grids}a). 

The Susskind fermion quasi-energy bandstructure \cite{Ham14a},
\begin{equation}
\cos^2 \varepsilon\delta t=(1-\gamma^2+\gamma^2\cos a_0k_x\cos a_0k_y)^2,\;\;\gamma\leq 1,\label{dispersioncos}
\end{equation}
has two inequivalent Dirac cones in the first Brillouin zone ${\cal B}$ shown in Fig.\ \ref{fig_grids}c, defined by
\begin{equation}
{\cal B}=\{k_x,k_y\in\mathbb{R}|-\pi/a_0<k_x,k_y\leq \pi/a_0\}.\label{BZdef}
\end{equation}
This is an improvement over the naive discretization, without staggering, which would have four inequivalent Dirac cones, at $(a_0k_x,a_0k_y)=(0,0)$, $(\pi,\pi)$, $(\pi,0)$, and $(0,\pi)$. Susskind fermions do not have the last two, but the first two Dirac cones remain.

In Fig.\ \ref{fig_grids}b,d we show the HPA modification of the staggered lattice discretization. Comparison with Fig.\ \ref{fig_grids}a,c shows that the HPA unit cell has one half the area of the unit cell of the original square lattice. Accordingly, the first Brillouin zone ${\cal B}'$, defined by
\begin{equation}
{\cal B}'=\{k_x,k_y\in\mathbb{R}|-2\pi/a_0<|k_x\pm k_y|\leq 2\pi/a_0\},\label{BZprimedef}
\end{equation}
has twice the area of ${\cal B}$.

\begin{figure}[tb]
\centerline{\includegraphics[width=0.7\linewidth]{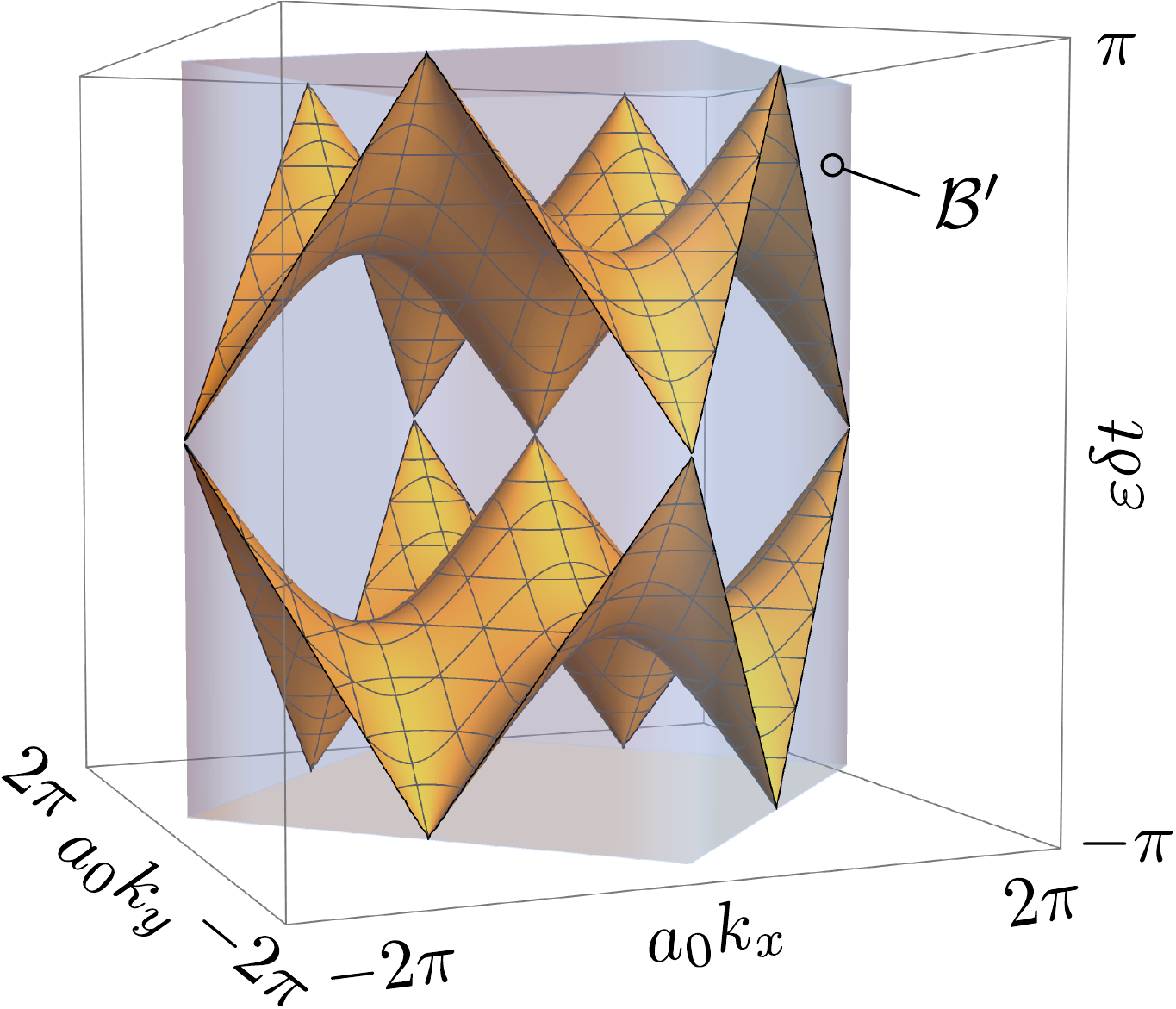}}
\caption{Quasi-energy bandstructure  \eqref{dispersionsin} of the HPA staggered lattice discretization, for $\gamma=1/\sqrt 2$, in the first Brillouin zone ${\cal B}'$ given by Eq.\ \eqref{BZprimedef}. There are two inequivalent Dirac cones, at center and corner of the Brillouin zone.
}
\label{fig_HPABZ}
\end{figure}

Inspection of the HPA dispersion \eqref{dispersionsin} then shows that, indeed, within ${\cal B}$ there is only a single Dirac cone, at $\bm{k}=0$. However, within ${\cal B}'$ there is a second cone at the corner $\bm{k}=(2\pi/a_0,0)$, see Fig.\ \ref{fig_HPABZ}. (The other Brillouin zone corners are related by a reciprocal lattice vector, so they are equivalent.) We conclude that, once we account for the Brillouin zone doubling, the HPA discretization still suffers from fermion doubling.

\section{Klein tunneling}
\label{sec_consequence}

The second Dirac cone at the corner of the Brillouin zone ${\cal B}'$ is at a relatively large momentum, so it will not play a role if the potentials are smooth: only momenta near $\bm{k}=0$ then matter and fermion doubling becomes irrelevant. But realistic disorder potentials may well vary on the scale of the lattice constant, and then fermion doubling has noticeable consequences. 

We investigate that here for Klein tunneling \cite{All11,Bee08}: Massless Dirac fermions are transmitted with unit probability when they approach a potential barrier at normal incidence, because conservation of chirality does not allow backscattering within a single Dirac cone. Coupling to a second cone will spoil that.

We contrast the numerical results following from the HPA staggered lattice technique \cite{Ham14} with those obtained using a manifestly single-cone discretization method \cite{Don22} --- a split-operator implementation  of the Stacey discretization \cite{Sta82,Two08,Lem21}. To make this paper selfcontained, both methods are summarized in App.\ \ref{sec_numerics}. Our numerical codes are available in a repository \cite{repository}.

We calculate the time dependence of a state $\Psi(x,y,t)$ incident along the $x$-axis on a rectangular barrier of height $V_0$ and width $50\,a_0$. The initial state is a Gaussian wave packet,
\begin{equation}
\Psi(x,y,0)=(4\pi w^2)^{-1/2}e^{ik_0x}e^{-(x^2+y^2)/2w^2}{1\choose 1},\label{wavepacket}
\end{equation}
with parameters $k_0=0.5/a_0$, $w=30\,a_0$, normalized such that $\int |\Psi|^2\,d\bm{r}=1$. We choose the time step $\delta t$ such that $\gamma=v_0\delta t/a_0=1/\sqrt 2$. The mean energy is $\bar{E}=\hbar v_0k_0=0.35\,\hbar/\delta t$, much less than the barrier height. The transmission probability $T$ is obtained from the integral of $|\Psi|^2$ over the area to the right of the barrier, at the late time $t=549\,\delta t$.  

\begin{figure}[tb]
\centerline{\includegraphics[width=1\linewidth]{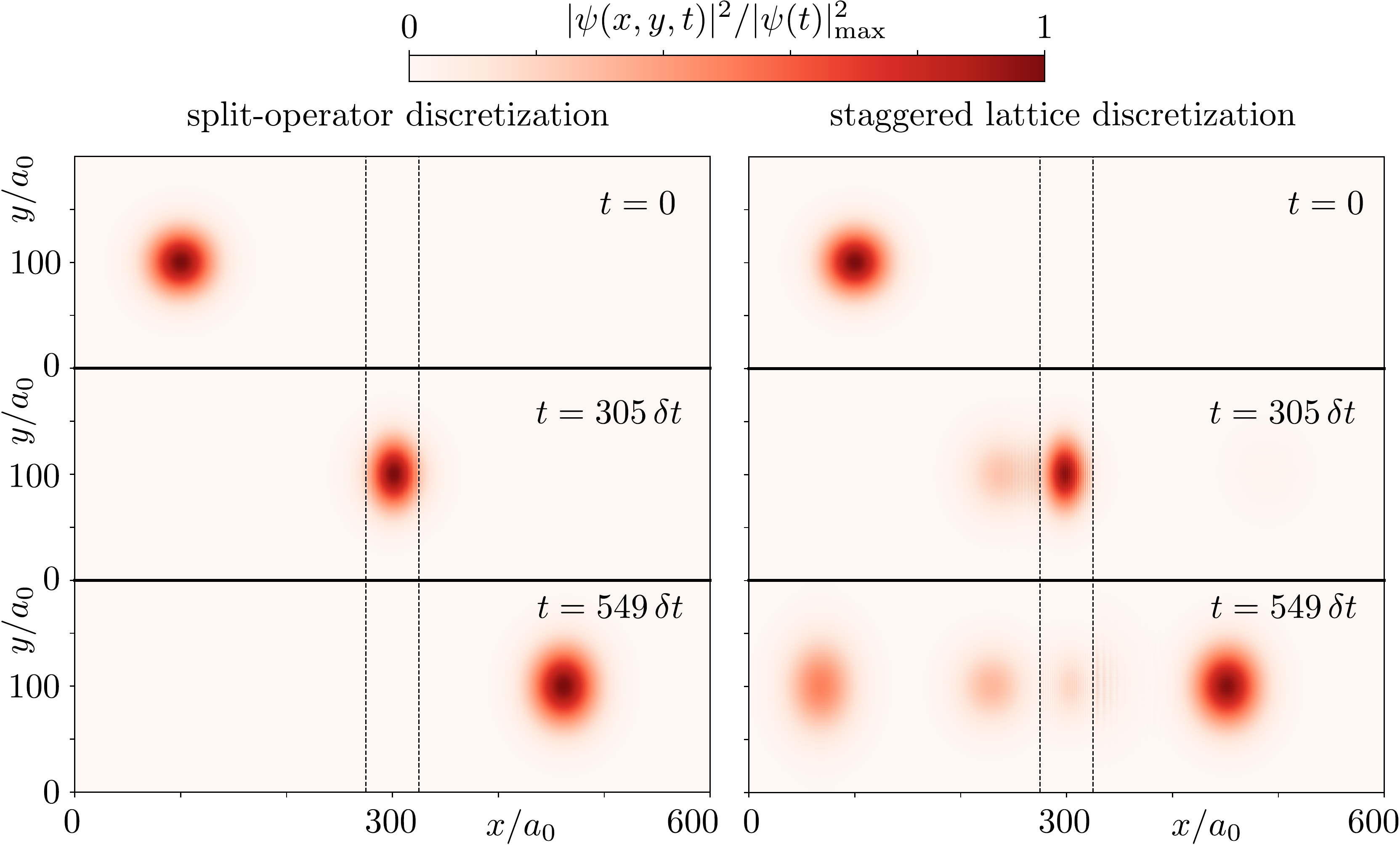}}
\caption{Three snapshots of the time-dependent simulation of Klein tunneling, in two alternative methods of discretization of the Dirac equation. A potential barrier of height $V_0=1.41\,\hbar/\delta t$ is located between the dotted lines. A wave packet at lower energy ($\bar{E}=0.35\,\hbar/\delta t$) is normally incident on the barrier. The color scale shows $|\Psi|^2$ normalized to unit peak height at each of the three times.
}
\label{fig_Klein1}
\end{figure}

\begin{figure}[tb]
\centerline{\includegraphics[width=0.7\linewidth]{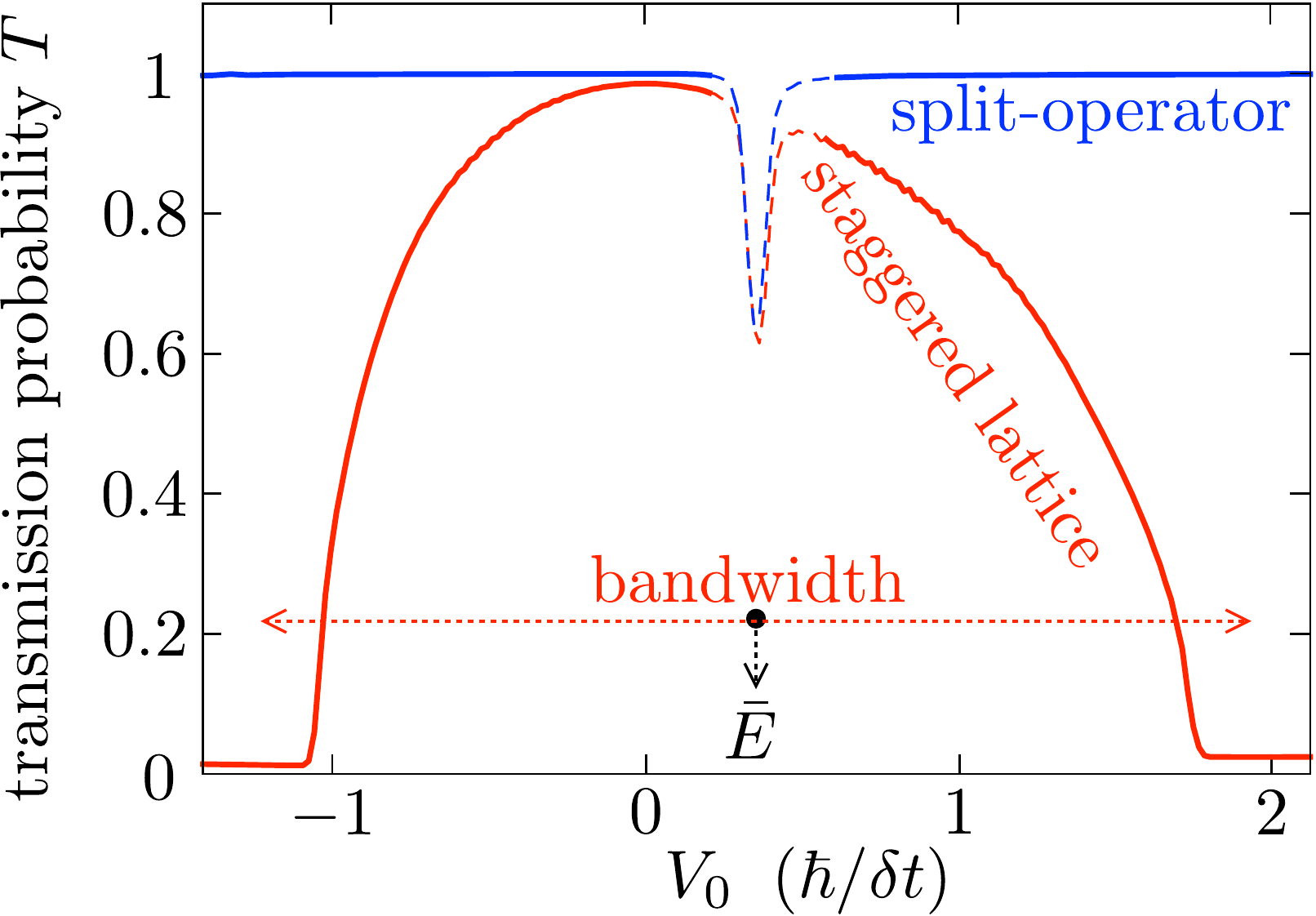}}
\caption{Transmission probability $T$ through the potential barrier of Fig.\ \ref{fig_Klein1}, as a function of the barrier height $V_0$. The blue and red curves are for, respectively, the split-operator discretization and the staggered lattice discretization. The mean energy $\bar{E}=0.35\,\hbar/\delta t$ of the incident wave packet \eqref{wavepacket} is indicated, as well as the finite bandwidth $\pi\hbar/\delta t$ of the staggered discretization. For the split-operator discretization $T\approx 1$ once $V_0\gtrsim\bar{E}$, while for the staggered discretization $T$ drops significantly below 1 well before $V_0-\bar{E}$ reaches the bandwidth.
}
\label{fig_transmission}
\end{figure}

As shown in Figs.\ \ref{fig_Klein1} and \ref{fig_transmission}, when $V_0$ is larger than $\bar{E}$ the wave packet is fully transmitted when the Dirac equation is discretized using the split-operator method, but not in the HPA staggered lattice discretization.\footnote{When $V_0$ is close to $\bar{E}$ the wave packet disperses side ways and backwards in the barrier region, hence the dashed dip in Fig.\ \ref{fig_transmission}. This is not a lattice artefact, the dip would also appear in the continuum description.} For example, when $V_0=2\bar{E}$ we find, respectively, $T=1.00$ and $T = 0.87$. We attribute the difference to fermion doubling.

\begin{figure}[tb]
\centerline{\includegraphics[width=0.7\linewidth]{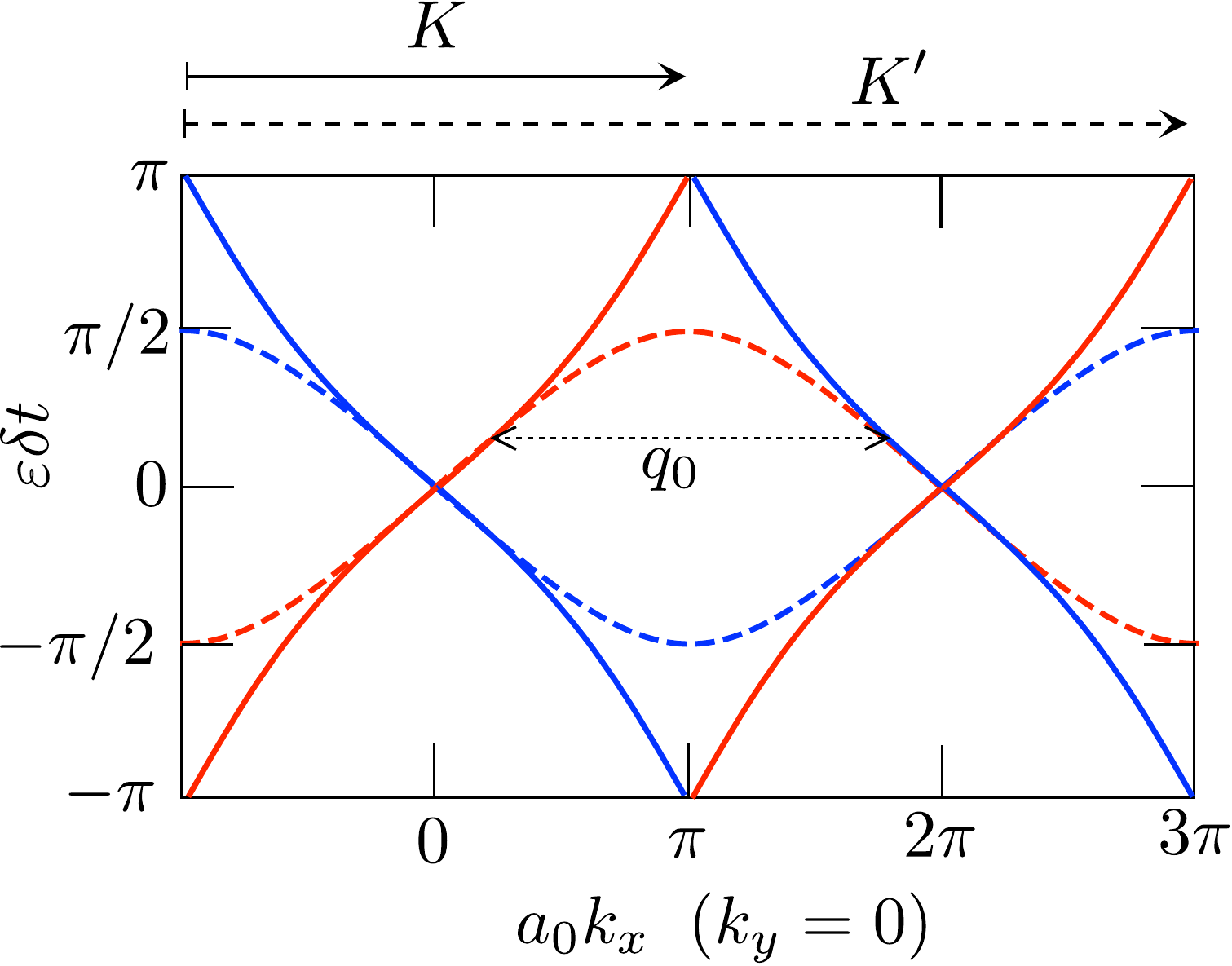}}
\caption{Dispersion relation along the $k_x$-axis for the split-operator discretization [solid curve, given by $\tan(\varepsilon\delta t/2)=\pm\gamma\tan(a_0k_x/2)$], and for the staggered lattice discretization [dashed curve, given by $\sin(\varepsilon\delta t/2)=\pm\gamma\sin(a_0k_x/2)$], both plotted for $\gamma=1/\sqrt 2$. The color red or blue distinguishes the eigenvalue $\pm 1$ of $\sigma_x$ (the chirality). The vectors $K$ and $K'$ are reciprocal lattice vectors for, respectively, the tangent and sine dispersions. A scalar potential can only couple branches of the same chirality. The momentum transfer $q_0$ thus leads to backscattering for the sine dispersion but it is forbidden for the tangent dispersion.
}
\label{fig_chirality}
\end{figure}

\begin{figure}[tb]
\centerline{\includegraphics[width=1\linewidth]{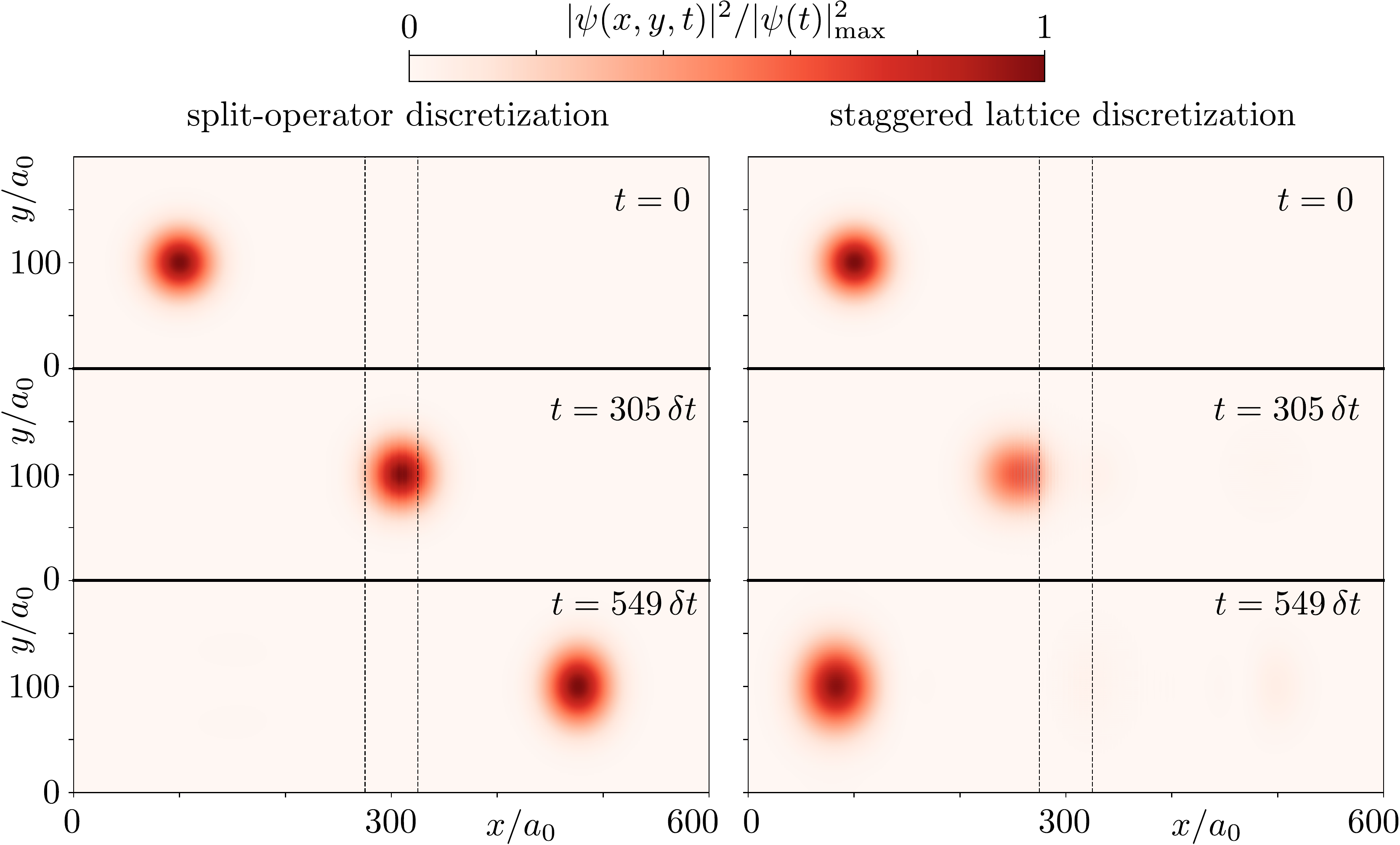}}
\caption{Same as Fig.\ \ref{fig_Klein1}, but for the modulated potential step \eqref{Vmodulated} (with parameters $V_0=0.71\,\hbar/\delta t$, $\delta V=0.071\,\hbar/\delta t$, $\bar{E}=0.35\,\hbar/\delta t$).
}
\label{fig_Klein2}
\end{figure}

To establish this, we have repeated the calculation with a periodic modulation of the barrier height,
\begin{equation}
V(x,y)=\begin{cases}
0&\text{if}\;\;|x/a_0-300|>25,\\
V_0+\delta V\sin q_0x&\text{if}\;\;|x/a_0-300|<25.
\end{cases}\label{Vmodulated}
\end{equation}
The wave number $q_0=2\pi/a_0-2k_0$ is chosen such that it couples a right-moving state at energy $\bar{E}=\hbar v_0k_0$ in the Dirac cone centered at $\bm{k}=(0,0)$ to a left-moving state in the Dirac cone centered at $\bm{k}=(2\pi/a_0,0)$. As explained in Fig.\ \ref{fig_chirality}, this coupling is forbidden by chirality conservation for the split-operator discretization, while it is allowed for the staggered lattice discretization. Fig.\ \ref{fig_Klein2} shows that, indeed, a small potential modulation causes a nearly complete suppression of the transmission ($T=0.06$) for the latter discretization only.

The suppressed transmission can be understood as the consequence of the opening of a gap at the Dirac point in the barrier region. The gapless Dirac cone is protected by time-reversal symmetry if there is only a single cone, but fermion doubling breaks that topological protection \cite{Kan13}. In App.\ \ref{app_gapopening} we calculate the bandstructure for the staggered lattice discretization in the presence of the periodic potential $V(x,y)=V_0\cos(2\pi x/a_0)$. Along the $k_y=0$ axis it is given by
\begin{equation}
\sin^2(\varepsilon\delta t/2)=\frac{(V_0\delta t/2)^2+\gamma^2\sin^2(a_0k_x/2)}{1+(V_0\delta t/2)^2}.\label{gappedband}
\end{equation}
The gap at $\bm{k}=0$ equals $2V_0$ for $V_0\delta t\ll 1$.

One might wonder at this stage whether the staggered lattice discretization is in any way an improvement over the naive discretization of the Dirac equation, without any staggering of the grid points. The staggering reduces the number of Dirac points in the 2D Brillouin zone from four to two ---  this is one advantage. But the coupling between the Dirac points is equally detrimental to Klein tunneling in the two discretization schemes, see App.\ \ref{app_naive}.

\section{Conclusion}
\label{sec_conclude}

In conclusion, we have uncovered a difficulty of staggered space-time lattice discretizations of the Dirac equation. In 2D staggered fermions \textit{a la} Susskind have two Dirac cones in the Brillouin zone \cite{Sus77}. To eliminate this lattice artefact known as fermion doubling, Hammer, P\"{o}tz, and Arnold \cite{Ham14} introduced a space-time lattice with bandstructure
\begin{equation}
\varepsilon=\pm 2\arcsin\left(v\sqrt{\sin^2 (k_x/2)+\sin^2 (k_y/2)}\right)\label{sindispersion}
\end{equation}
(in units where $a_0$ and $\delta t$ are 1). The Susskind fermion Brillouin zone is $-\pi<k_x,k_y<\pi$ and in that Brillouin zone the bandstructure \eqref{sindispersion} has only a Dirac cone at the origin $\bm{k}=0$. 

What we have found is that this bandstructure is accompanied by Brillouin zone doubling: Along the $k_x$-axis it extends from $-2\pi<k_x<2\pi$, so the Dirac cone at $\bm{k}=(2\pi,0)$ is independent from the one at the origin --- they are not related by a reciprocal lattice vector. We have shown that this fermion doubling has physical consequences in the breakdown of Klein tunneling.

To ascertain that fermion doubling is at the origin of these effects, we have compared with an alternative space-time discretization using a split-operator technique \cite{Don22}, with bandstructure
\begin{equation}
\varepsilon=\pm 2\arctan\left(v\sqrt{\tan^2 (k_x/2)+\tan^2 (k_y/2)}\right).\label{tandispersion}
\end{equation}
The replacement of sine by tangent avoids the Brillouin zone doubling, essentially because $\sin(k/2)$ is $4\pi$-periodic in $k$, while $\tan(k/2)$ is $2\pi$-periodic. The Dirac cones at $\bm{k}=0$ and $\bm{k}=(2\pi,0)$ are now equivalent, related by a reciprocal lattice vector, and indeed we recover the Klein tunneling with unit probability expected for massless Dirac fermions.

The staggered lattice discretization has one feature that the split-operator discretization lacks: the possibility to include the vector potential in a fully gauge invariant way via the Peierls substitution \cite{Ham14,Pot17a}. We are inclined to think that this is an intrinsic limitation of single-cone discretization schemes, but we have not succeeded in deriving a ``no-go'' theorem that forbids gauge invariance without fermion doubling.

\acknowledgments

This project has received funding from the Netherlands Organization for Scientific Research (NWO/OCW) and from the European Research Council (ERC) under the European Union's Horizon 2020 research and innovation programme. 

We acknowledge correspondence on our preprint with R. Hammer and W. P\"{o}tz, which has led to a corrigendum at J. Comp.\ Phys.\ \textbf{457}, 111118 (2022).

\appendix

\section{Two methods of space-time discretization of the Dirac equation}
\label{sec_numerics}

In the main text we compare results from two space-time lattice discretizations of the Dirac equation, the staggered lattice approach of Ref.\ \onlinecite{Ham14} and the split-operator approach of Ref.\ \onlinecite{Don22}. We summarize these two methods.

\subsection{Staggered lattice approach}
\label{sec_staggeredapproach}

\begin{figure}[tb]
\centerline{\includegraphics[width=0.7\linewidth]{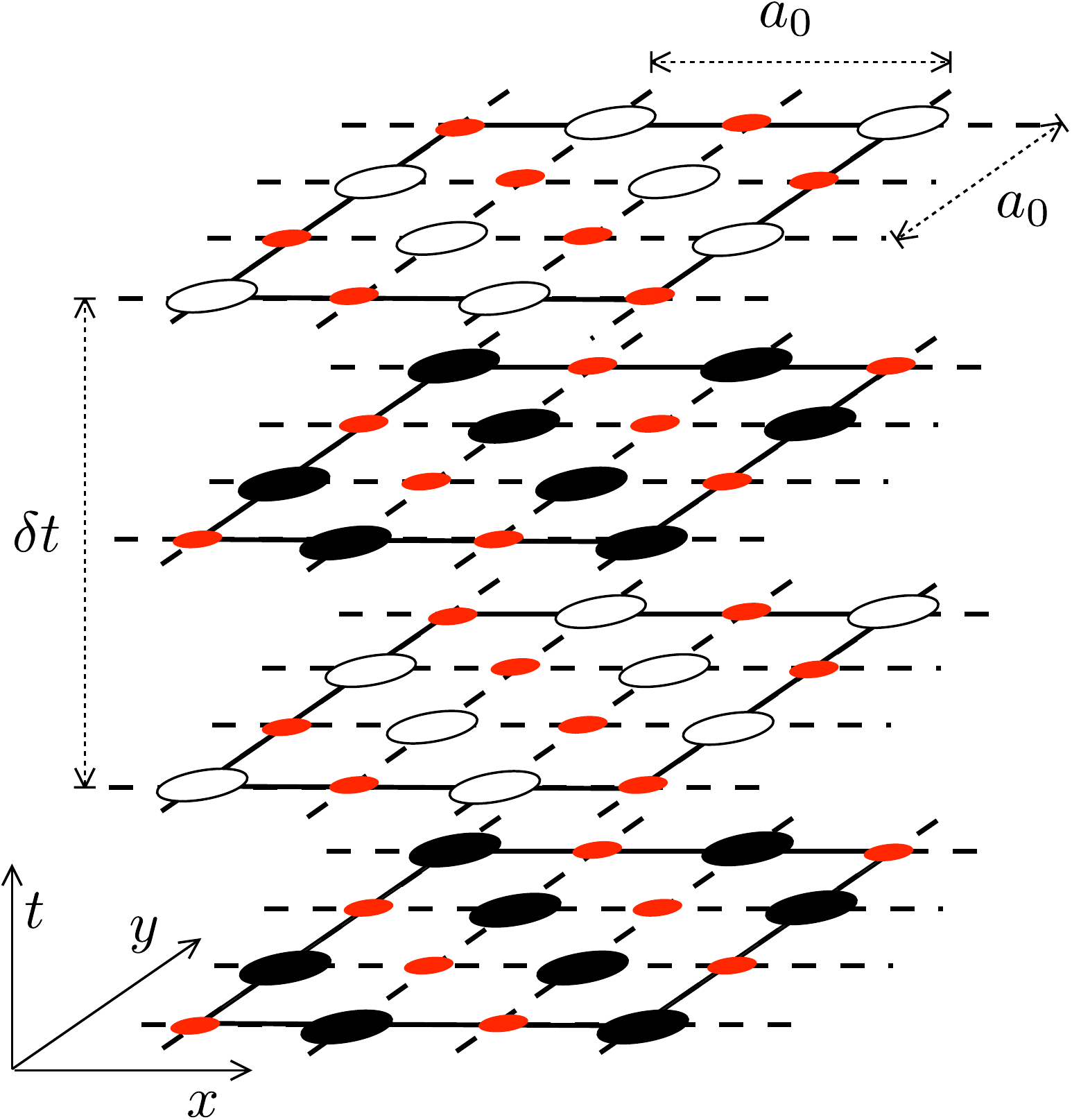}}
\caption{Space-time lattice in the HPA method of staggered lattice discretization of the $2+1$ dimensional Dirac equation \cite{Ham14}. The $u$ and $v$ components of the spinor wave function $\Psi=(u,v)$ are indicated by black and white dots, respectively. The finite differences are evaluated at the red points.
}
\label{fig_HPA}
\end{figure}

Hammer, P\"{o}tz, and Arnold \cite{Ham14} discretize the $2+1$ dimensional Dirac equation on the space-time lattice shown in Fig.\ \ref{fig_HPA}. The two components of the wave function $\Psi=(u,v)$ are evaluated on two different lattices, staggered in both space and time. The $v$-lattice is obtained from the $u$-lattice by a translation of $\delta t/2$ in the time direction and by $a_0/2$ in the $x$-direction. A translation of either $u$-lattice or $v$-lattice by $a_0/2$ in the $x$-direction without a time translation defines a third lattice of points $\bm{S}_{nms}=(x_n,y_m,t_s)$, the red points in Fig.\ \ref{fig_HPA}. Each of these three lattices is face-centered square in the $x$--$y$ plane, with the unit cell and Brillouin zone ${\cal B}'$ of Fig.\ \ref{fig_grids}b,d.

The finite-difference equation for the $u$ component is (abbreviating $\gamma=v_0\delta t/a_0$)
\begin{widetext}
\begin{subequations}
\label{HPAeqs}
\begin{align}
&i[u(x_n,y_m,t_s+\tfrac{1}{2}\delta t)-u(x_n,y_m,t_s-\tfrac{1}{2}\delta t)]=-i\gamma[v(x_n+\tfrac{1}{2}a_0,y_m,t_s)-v(x_n-\tfrac{1}{2}a_0,y_m,t_s)]\nonumber\\
&\quad-\gamma[v(x_n,y_m+\tfrac{1}{2}a_0,t_s)-v(x_n,y_m-\tfrac{1}{2}a_0,t_s)]+\frac{\delta t}{2\hbar}V(x_n,y_m,t_s)[u(x_n,y_m,t_s+\tfrac{1}{2}\delta t)+u(x_n,y_m,t_s-\tfrac{1}{2}\delta t)],
\end{align}
for $(x_n,y_m,t_s\pm \tfrac{1}{2}\delta t)$ on the $u$-lattice. The arguments of the $v$-component are then located on the $v$-lattice. Similarly, the finite-difference equation for the $v$-component is
\begin{align}
&i[v(x_n,y_m,t_s+\tfrac{1}{2}\delta t)-v(x_n,y_m,t_s-\tfrac{1}{2}\delta t)]=-i\gamma[u(x_n+\tfrac{1}{2}a_0,y_m,t_s)-u(x_n-\tfrac{1}{2}a_0,y_m,t_s)]\nonumber\\
&\quad+\gamma[u(x_n,y_m+\tfrac{1}{2}a_0,t_s)-u(x_n,y_m-\tfrac{1}{2}a_0,t_s)]+\frac{\delta t}{2\hbar}V(x_n,y_m,t_s)[v(x_n,y_m,t_s+\tfrac{1}{2}\delta t)+v(x_n,y_m,t_s-\tfrac{1}{2}\delta t)],
\end{align}
\end{subequations}
\end{widetext}
for $(x_n,y_m,t_s\pm \tfrac{1}{2}\delta t)$ on the $v$-lattice. The computational cost of the solution of these difference equations scales linearly in $N$ on an $N$-site lattice.

The quasi-energy bandstructure for $V=0$ is given by
\begin{equation}
\sin^2(\varepsilon\delta t/2)=\gamma^2[\sin^2(a_0k_x/2)+\sin^2(a_0k_y/2)].\label{app-dispersionsin}
\end{equation}
The requirement of a real quasi-energy $\varepsilon$ restricts $\gamma^2\leq 1/2$. The bandstructure in the first Brillouin zone is plotted in Fig.\ \ref{fig_HPABZ}, for $\gamma=1/\sqrt 2$.

Figs.\ \ref{fig_Klein1} and \ref{fig_Klein2} show at time slice $t_s$ both $|v(x_n,y_m,t_s)|^2$ and $|u(x_n+1/2,y_m,t_s+1/2)|^2$, each on its own staggered lattice. Because these amplitudes vary little over a lattice spacing other ways to compute $|\Psi|^2$, by averaging over nearby sites \cite{Ham14}, do not make a significant difference. 

\subsection{Split-operator approach}
\label{sec_splitoperator}

\begin{figure}[tb]
\centerline{\includegraphics[width=0.7\linewidth]{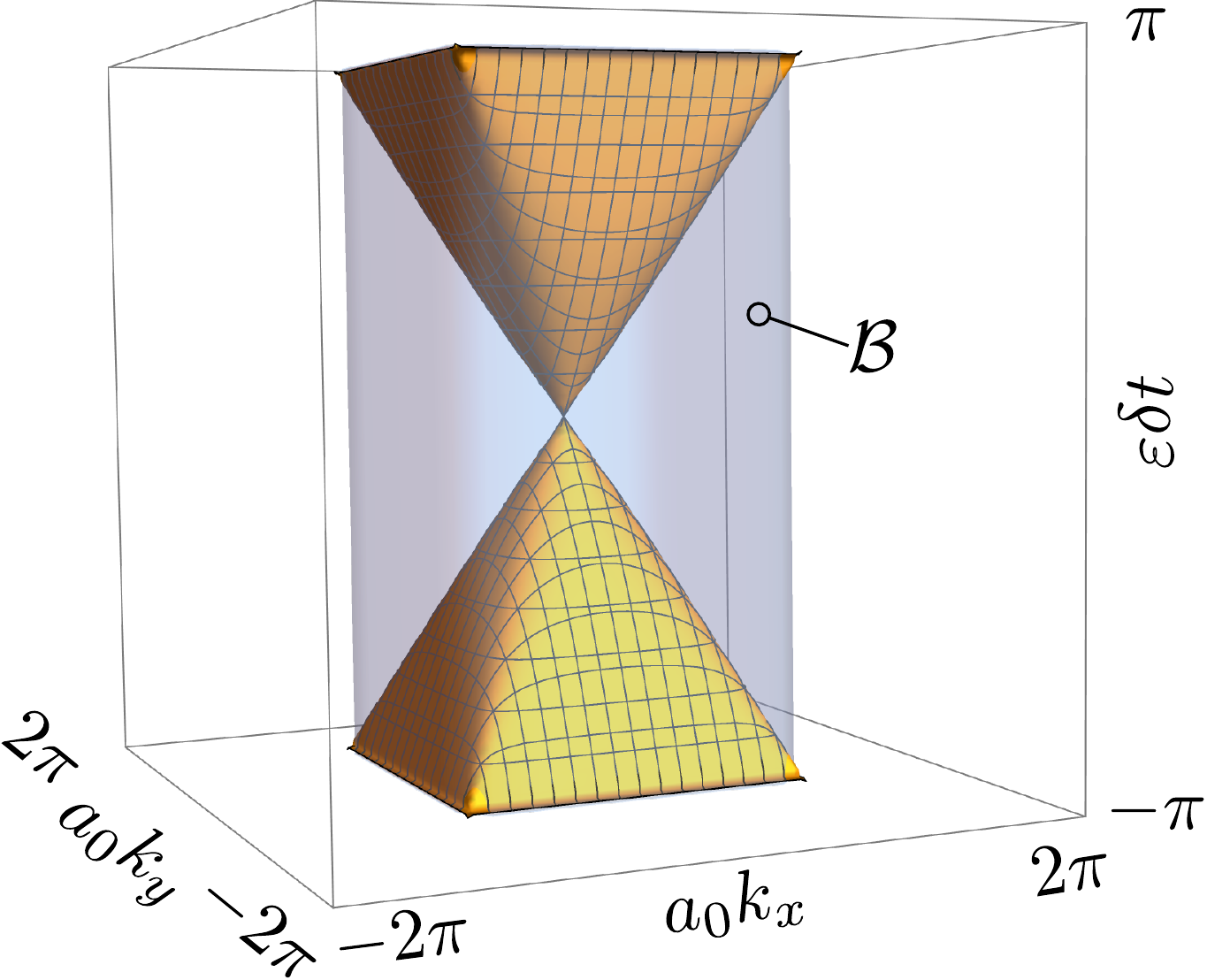}}
\caption{Quasi-energy bandstructure \eqref{app-dispersiontan}  of the split-operator discretization, for $\gamma=1/\sqrt 2$, in the first Brillouin zone ${\cal B}$ given by Eq.\ \eqref{BZdef}. There is only a single Dirac cone, at the center of the Brillouin zone.
}
\label{fig_splitBZ}
\end{figure}

The split-operator approach of Ref.\ \onlinecite{Don22} uses the same regular square lattice for both $u$ and $v$ components (Brillouin zone $|k_x|,|k_y|<\pi/a_0$). The time evolution $\Psi(t+\delta t)=U\Psi(t)$ is given by the unitary operator product (``split operator'')
\begin{align}
U={}&e^{-iV(\bm{r})\delta t/2\hbar}{\cal F}^{-1}\frac{1-i\gamma\,\sum_\alpha \sigma_\alpha\tan(a_0k_\alpha/2)}{1+i\gamma\,\sum_\alpha \sigma_\alpha\tan(a_0k_\alpha/2)}\nonumber\\
&\cdot{\cal F}e^{-iV(\bm{r})\delta t/2\hbar}.\label{Udef}
\end{align}
The Fourier transform ${\cal F}$ performs a change of basis, so that the $\bm{r}$-dependent operators are evaluated in the real-space basis and the $\bm{k}$-dependent operators are evaluated in the momentum basis --- at minimal computational cost. The cost of a Fast Fourier Transform scales as $N\log N$ on an $N$-site lattice.

The eigenvalues $e^{i\varepsilon t}$ of $U$ for $V=0$ depend on $\bm{k}$ according to
\begin{equation}
\tan^2(\varepsilon\delta t/2)=\gamma^2[\tan^2(a_0k_x/2)+\tan^2(a_0k_y/2)].\label{app-dispersiontan}
\end{equation}
The quasi-energy $\varepsilon$ is real for any $\gamma>0$. The bandstructure in the first Brillouin zone is plotted in Fig.\ \ref{fig_splitBZ}, for $\gamma=1/\sqrt 2$.

\section{Gap opening for the staggered lattice discretization}
\label{app_gapopening}

Because the staggered lattice discretization has two Dirac cones in the Brillouin zone, the gapless Dirac point is not protected by time-reversal symmetry --- a gap can open without violating Kramers degeneracy. Here we show this by an explicit calculation.

The gap opening mechanism can be described as ``fold and split'': a potential that varies on the scale of the lattice constant $a_0$ folds the Dirac cone at $\bm{k}=(2\pi/a_0,0)$ onto the cone at $\bm{k}=(0,0)$, and then the upper and lower cone can split apart while preserving the double degeneracy required by Kramers theorem.

We consider the periodic potential $V(x,y)=V(x+a_0,y)=V(x,y+a_0)$ and solve the finite difference equations \eqref{HPAeqs} for the Bloch state $\Psi(x+a_0,y,t)=e^{ia_0k_x}\Psi(x,y,t)$, $\Psi(x,y+a_0,t)=e^{ia_0k_y}\Psi(x,y,t)$. There are four independent equations, involving the spinor amplitudes
\begin{align}
&u_1(t)=u(0,0,t-\delta t/2),\;\;
u_2(t)=u(a_0/2,a_0/2,t-\delta t/2),\nonumber\\
&v_1(t)=v(a_0/2,0,t),\;\;v_2(t)=v(0,a_0/2,t),
\end{align}
and potential values $V_A=V(0,0)$, $V_B=V(a_0/2,a_0/2)$, $V_C=V(a_0/2,0)$, $V_D=V(0,a_0/2)$. The four equations can be written in the matrix form
\begin{widetext}
\begin{subequations}
\begin{align}
&{\cal P}\begin{pmatrix}
    u_1(t+\delta t)\\u_2(t+\delta t)\\
    v_1(t+\delta t)\\v_2(t+\delta t)
    \end{pmatrix}={\cal Q} \begin{pmatrix}
    u_1(t)\\u_2(t)\\
    v_1(t)\\v_2(t)
    \end{pmatrix},\;\; {\cal P}=  \begin{pmatrix}
    i/\delta t-V_A/2&0&0&0\\
    0&i/\delta t-V_B/2&0&0\\
    i(e^{ia_0 k_x}-1)&e^{-ia_0 k_y}-1&i/\delta t-V_C/2&0\\
    1-e^{ia_0 k_y}&i(1-e^{-ia_0 k_x})&0&i/\delta t-V_D/2\\
    \end{pmatrix},\\
& {\cal Q}=\begin{pmatrix}
    i/\delta t+V_A/2&0&i(e^{-ia_0 k_x}-1)&e^{-ia_0 k_y}-1\\
    0&i/\delta t+V_B/2&1-e^{ia_0 k_y}&i(1-e^{ia_0 k_x})\\
    0&0&i/\delta t+V_C/2&0\\
    0&0&0&i/\delta t+V_D/2\\
    \end{pmatrix}.
\end{align}
\end{subequations}
\end{widetext}

The eigenvalues $e^{i\varepsilon\delta t}$ of the matrix product ${\cal P}^{-1}{\cal Q}$ give the bandstructure $\varepsilon(k_x,k_y)$. One readily recovers Eq.\ \eqref{dispersionsin} for $V(x,y)\equiv 0$. For the potential $V(x,y)=V_0\cos(2\pi x/a_0)$ we set $V_A=V_D=V_0$, $V_B=V_D=-V_0$. Along the line $k_y=0$ we then find the result \eqref{gappedband}, with a gap at $\bm{k}=0$ of size
\begin{equation}
\Delta\varepsilon=\frac{4}{\delta t}\arcsin\left(\frac{V_0\delta t/2}{\sqrt{1+(V_0\delta t/2)^2}}\right).
\end{equation}

We note that the topological protection of the Dirac cone for the split-operator discretization \eqref{Udef} was established in Ref.\ \onlinecite{Don22}.

\section{Klein tunneling of naive fermions}
\label{app_naive}

Fig.\ \ref{fig_transmission} compares the Klein tunneling probability for staggered-lattice and split-operator discretizations. For completeness, here we compare to the naive discretization, without any staggering. 

We discretize the Dirac equation \eqref{Diraceq} on a space-time lattice by means of the Crank-Nicolson method,
\begin{align}
&\left(1-\frac{i\delta t}{2\hbar}H\right)\Psi(\bm{r},t+\delta t)=\left(1+\frac{i\delta t}{2\hbar} H\right)\Psi(\bm{r},t),\\
&H\Psi(\bm{r},t)=V(\bm{r})\Psi(\bm{r},t)\nonumber\\
&\;+\frac{\hbar v_0}{2ia_0}\sum_{\alpha=x,y}\sigma_\alpha[\Psi(\bm{r}+a_0\bm{\hat r}_\alpha,t)-\Psi(\bm{r}-a_0\bm{\hat r}_\alpha,t)].
\end{align}
The unit vectors $\bm{\hat r}_x,\bm{\hat r}_y$ point in the $x$- and $y$-directions. The vector potential may be included by Peierls substitution, but here we take zero magnetic field.

The naive-fermion bandstructure
\begin{equation}
\tan^2(\epsilon \delta t/2)=\tfrac{1}{4}\gamma^2(\sin^2 a_0k_x+\sin^2 a_0k_y),\;\;\gamma=v_0\delta t/a_0,
\end{equation}
has four inequivalent Dirac points in the first Brillouin zone, at $a_0\bm{k}=(0,0),(0,\pi),(\pi,0)$, and $(\pi,\pi)$. The staggered discretization reduces that to two Dirac points. 

The naive-fermion band width for motion in the $x$-direction is $(4\hbar/\delta t)\arctan(\gamma/2)$. For the same lattice constants this is smaller than the band width $(4\hbar/\delta t)\arcsin\gamma$ of the staggered discretization --- as expected, because the staggering introduces additional lattice points in the unit cell (see Fig.\ \ref{fig_grids}). To compare the two discretization schemes at the same band width, we take $\gamma=1/\sqrt 2$ for the staggered discretization and $\gamma=2$ for the naive discretization --- then in both cases the band width is $\pi\hbar/\delta t$.

\begin{figure}[tb]
\centerline{\includegraphics[width=0.7\linewidth]{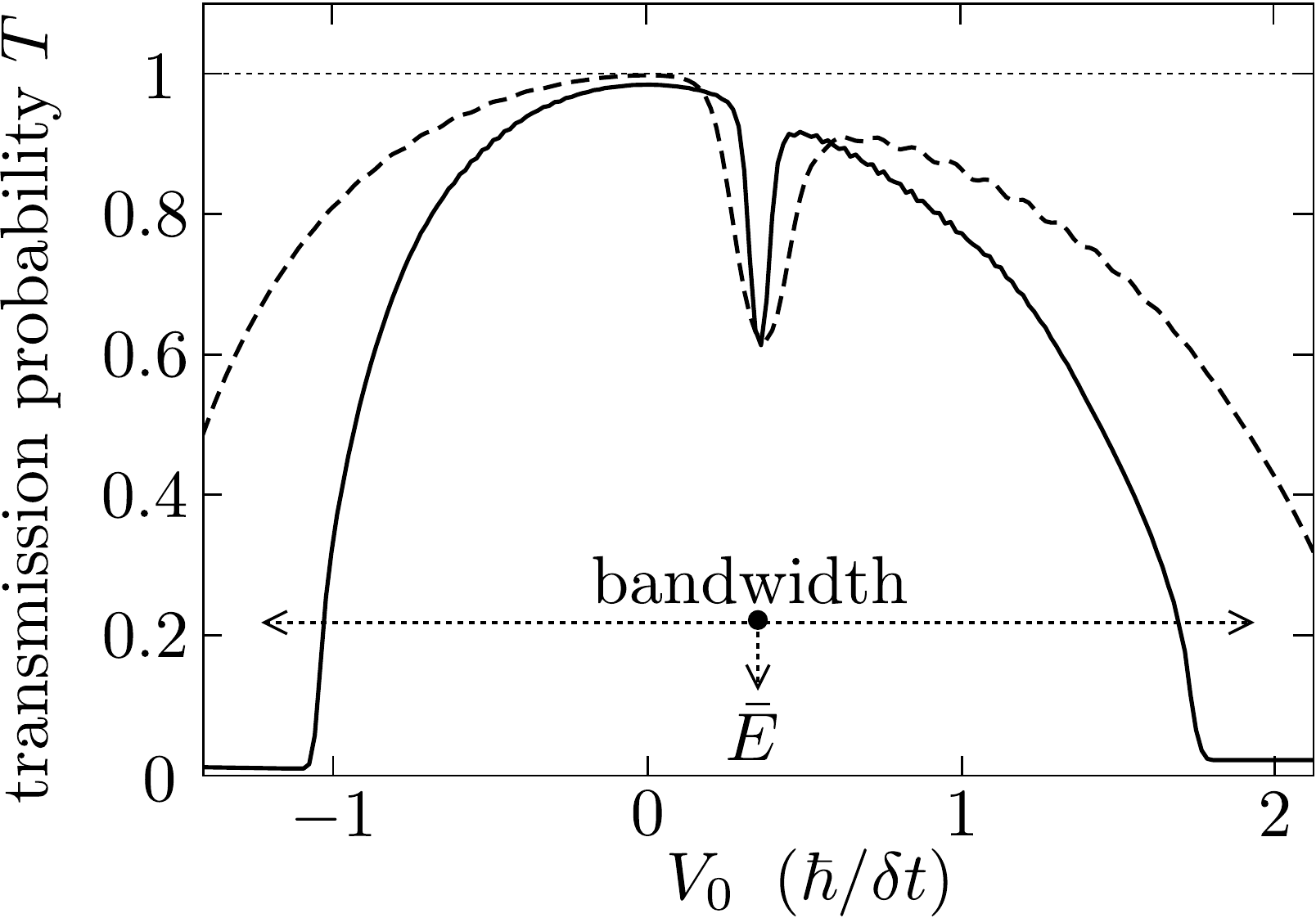}}
\caption{Same as Fig.\ \ref{fig_transmission}, but now comparing the staggered-fermion discretization (solid curve) with the naive discretization (dashed curve). To have the same band width $\pi\hbar/\delta t$ in both cases we rescaled $v_0$ such that $\gamma=1/\sqrt 2$ in the former case and $\gamma=2$ in the latter case.
}
\label{fig_naive}
\end{figure}

Results are shown in Fig.\ \ref{fig_naive}. We conclude that the staggering does not significantly improve the Klein tunneling.

 \end{document}